\documentclass{PoS}

\usepackage{amsmath}
\usepackage{graphicx}

\title{Strangeness neutral equation of state with a critical point}

\ShortTitle{Strangeness neutral equation of state with a critical point}

\author{\speaker{Jamie M. Karthein}\\
        Department of Physics, University of Houston, Houston, TX 77204, USA\\
        E-mail: \email{jamie@karthein.com}}

\author{Debora Mroczek\\
        Department of Physics, 
University of Illinois at Urbana-Champaign, Urbana, IL 61801, USA}

\author{Angel R. Nava Acuna\\
        Department of Physics, University of Houston, Houston, TX 77204, USA}  

\author{Jacquelyn Noronha-Hostler\\
        Department of Physics, 
University of Illinois at Urbana-Champaign, Urbana, IL 61801, USA}          

\author{Paolo Parotto\\
        University of Wuppertal, Department of Physics, Wuppertal D-42119, Germany}             
     
\author{Damien R. P. Price\\
        Department of Physics, University of Houston, Houston, TX 77204, USA}

\author{Claudia Ratti\\
        Department of Physics, University of Houston, Houston, TX 77204, USA}

\abstract{We formulate a family of equations of state for Quantum Chromodynamics that exhibit critical features and obey the charge conservation conditions present in heavy-ion collisions (HICs). This construction utilizes the first-principle Lattice QCD (LQCD) results up to $\mathcal{O}(\mu_B^4)$ by matching the Taylor coefficients at each order. The criticality of the equation of state (EoS) is implemented based on the well-established principle of universal scaling behavior, where QCD belongs to the 3D Ising Model universality class. The critical point can be placed in a range of temperature and baryonic chemical potential relevant for the Beam Energy Scan II at RHIC. Furthermore, the strength of the critical features can be varied as well. This flexibility is embedded in four free parameters, that could potentially be constrained by the experimental data. We will discuss the features and versatility of our EoS, as well as present the relevant thermodynamic quantities which are important for hydrodynamical simulations of HICs. }

\FullConference{International Conference on Critical Point and Onset of Deconfinement\\
                March 15 - 19, 2021\\
                Zoom}

\begin{document}

\section{Introduction}
The equation of state (EoS) for Quantum Chromodynamics (QCD) plays a crucial role in the search for a proposed critical endpoint in the QCD phase diagram. The EoS is needed as input in the hydrodynamic simulations that describe the system created during HICs \cite{Dexheimer:2020zzs,Monnai:2021kgu}. The determination of the QCD equation of state and phase diagram at finite density from first principles is currently limited,  due to the Fermi sign problem. 
To support the RHIC experimental effort, theoretical predictions on the location of the critical point \cite{Critelli:2017oub,Bazavov:2017dus,DElia:2016jqh,Pasztor:2018yae,Fodor:2004nz,Fischer:2012vc,Borsanyi:2020fev, Fu:2019hdw} and its effect on observables \cite{An:2020vri,Mroczek:2020rpm,Bluhm:2020mpc} have recently been developed and explored.  
Along those lines,  a family of equations of state, which agree with the one from lattice QCD in the density regime where it is available,as well as including a critical point in the 3D Ising model universality class (the one expected for QCD) has been developed within the framework of the BEST collaboration \cite{Parotto:2018pwx,Karthein:2021nxe}.
This approach provides an EoS in a region of the phase diagram relevant for Beam Energy Scan physics. Furthermore, the location of the critical point and the strength of the critical region can be chosen by the user.  This approach can thus be constrained through comparisons with the experimental data from Beam Energy Scan II.  This EoS has already been in use for the study of out-of-equilibrium approaches to the QCD critical point \cite{Dore:2020jye,Hama:2020nfp}, the sign of the kurtosis \cite{Mroczek:2020rpm}, and to calculate transport coefficients \cite{McLaughlin:2021dph}. 

\section{Methodology}
\label{method}
In order to study the effect of a critical point on QCD thermodynamics, we utilized the principle of universality classes to map the critical behavior onto the phase diagram of QCD. The 3D Ising model is the correct choice for this approach because it exhibits the same scaling features in the vicinity of a critical point as QCD, which is to say that they belong to the same universality class \cite{Pisarski:1983ms,Rajagopal:1992qz}. We implemented the non-universal mapping of the 3D Ising model onto the QCD phase diagram in such a way that the final pressure agrees with Lattice QCD results on the EoS where they are available. This prescription can be summarized as follows:
\begin{enumerate}
    \item Parametrize the 3D Ising model in the vicinity of the critical point, as previously done in the literature  \cite{Guida:1996ep, Nonaka:2004pg, Stephanov:2011pb, Parotto:2018pwx}, 
    \begin{equation}
     \label{IsingEoS}
        \begin{split}
            M &= M_0 R^{\beta} \theta \\
            h &= h_0 R^{\beta \delta} \tilde{h}(\theta) \\
            r &= R(1- \theta^2)
        \end{split}
    \end{equation}
	where the magnetization $M$, the  magnetic field $h$, and the reduced temperature $r$, are functions of the external parameters $R$ and $\theta$. The normalization constants for the magnetization and magnetic field are $M_0=0.605$ and $h_0=0.364$, respectively, $\beta=0.326$ and $\delta=4.8$ are critical exponents in the 3D Ising Model, and $\tilde{h}(\theta)=\theta (1-0.76201\theta^2+0.00804\theta^4$).
	
	This renormalization group approach leads to the following definition of the singular part of the Gibbs' free energy, from which one can get the pressure via a sign change:
	\begin{equation}
	\label{GFreeEner}
     \begin{split}
	     P_{\text{Ising}} &= - G(R,\theta) \\
	     &= h_0 M_0 R^{2 - \alpha}(\theta \tilde{h}(\theta) - g(\theta)),
	    \end{split}
	\end{equation}
	where $\tilde{h}(\theta)$ is defined above and $g(\theta)$ is a polynomial in $\theta$, the full form of which can be found in \cite{Parotto:2018pwx,Karthein:2021nxe}.
    \item Map the critical point onto the QCD phase diagram via a linear map from \{$T$, $\mu_B$\} to \{$r,h$\} at a location $\{T_c,\mu_{B,c}\}$: 
    \begin{equation} \label{mapT}
        \frac{T-T_c}{T_c}=\omega(\rho r \sin{\alpha_1} + h \sin{\alpha_2})
    \end{equation}
    \begin{equation} \label{mapmuB}
        \frac{\mu_B-\mu_{B,c}}{T_c} = \omega(-\rho r \cos{\alpha_1} - h \cos{\alpha_2})
    \end{equation}
    
    where $\alpha_1$ and $\alpha_2$ are the angles between the axes of the QCD phase diagram and the Ising model ones and $\omega$ and $\rho$ are scaling parameters for the Ising-to-QCD map.

    \item Reduce the number of free parameters in the non-universal mapping from six to four, by placing the critical point along the chiral phase transition line, such that the $r$ axis of the Ising model is tangent to the transition line of QCD at the critical point: 
    \begin{equation} \label{chiraltrans}
        T=T_0 + \kappa \, T_0 \, \left(\frac{\mu_B}{T_0}\right)^2 + \mathcal{O}(\mu_B^4).
    \end{equation}
    The choice we made for the critical point here was \{$T_c$=143.2 MeV, $\mu_{B,c}$=350 MeV\}, with angular parameters $\alpha_1$=3.85\textdegree and $\alpha_2$=93.85\textdegree, and  scaling parameters $\omega$=1 and $\rho$=2.  It is important to note that such a choice of parameters only has an illustrative purpose. We do not, by making this choice, claim any prediction for the location of the critical point or the size of the critical region. Here, we seek to provide an estimate of the effect of critical features on heavy-ion-collision systems,  and leave the freedom of choice of the parameters to the users. An additional comment regarding the scaling parameters is in order: varying the parameters $\omega$ and $\rho$ will increase or decrease the effects of the critical point \cite{Parotto:2018pwx,Bzdak:2019pkr,Mroczek:2020rpm}.  In order to reduce the parameter space, we are relying on experimental data from the BES-II to constrain the parameters and narrow down the location of the critical point. 

    \item Calculate the Ising model susceptibilities and perform the matching of the Taylor expansion coefficients order by order to Lattice QCD results at $\mu_B=0$. 
    This yields a non-Ising, or background, pressure, which is, by construction, the difference between the lattice and Ising contributions:
    \begin{equation} \label{coeffmatch}
        T^4 c_n^{\rm{LAT}}(T) = T^4 c_n^{\rm{Non-Ising}}(T) + T_c^4 c_n^{\rm{Ising}}(T)
    \end{equation}
\begin{figure*}[t]
    \centering
    \begin{tabular}{c c c}
    \includegraphics[width=0.33\textwidth]{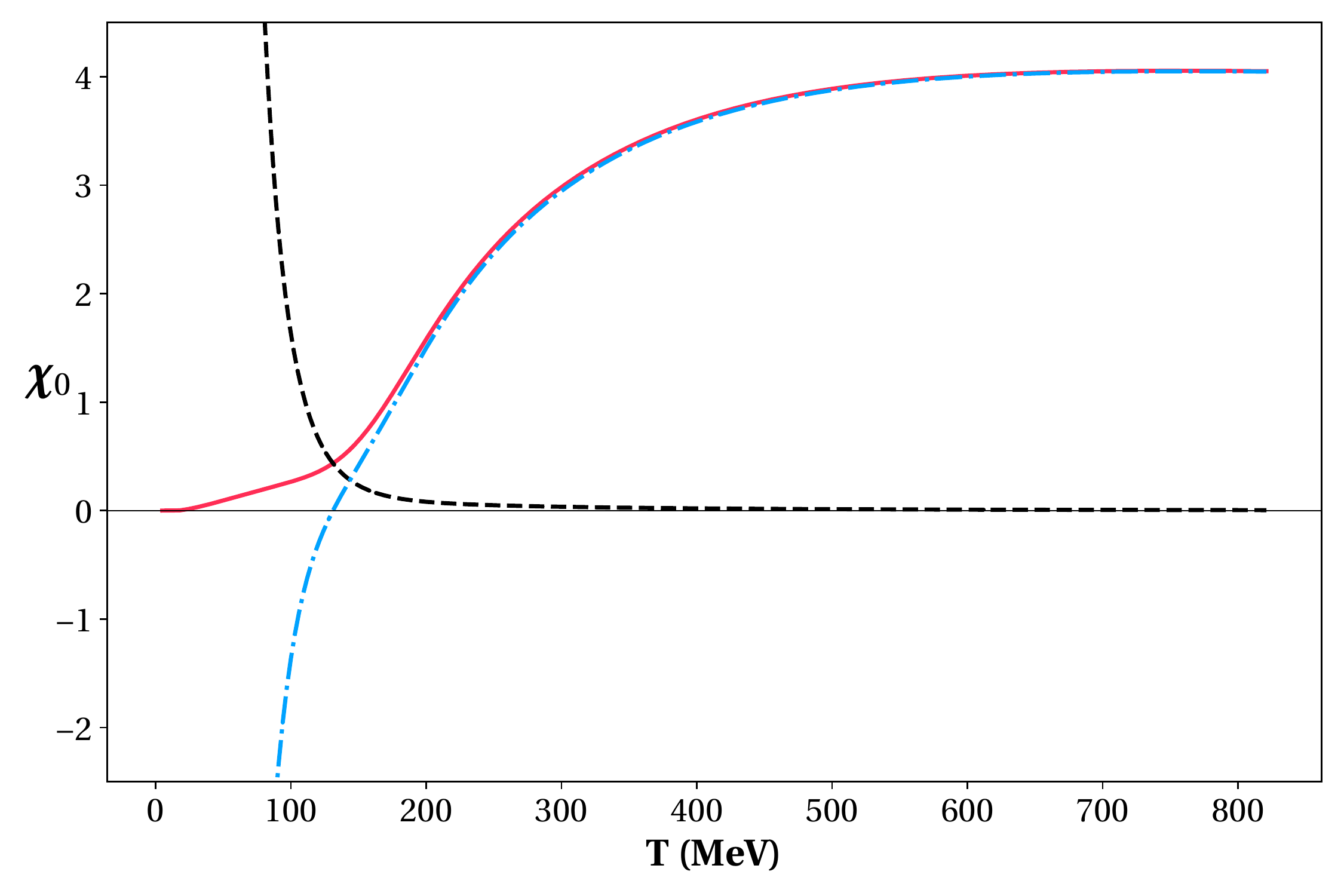} &
    \includegraphics[width=0.33\textwidth]{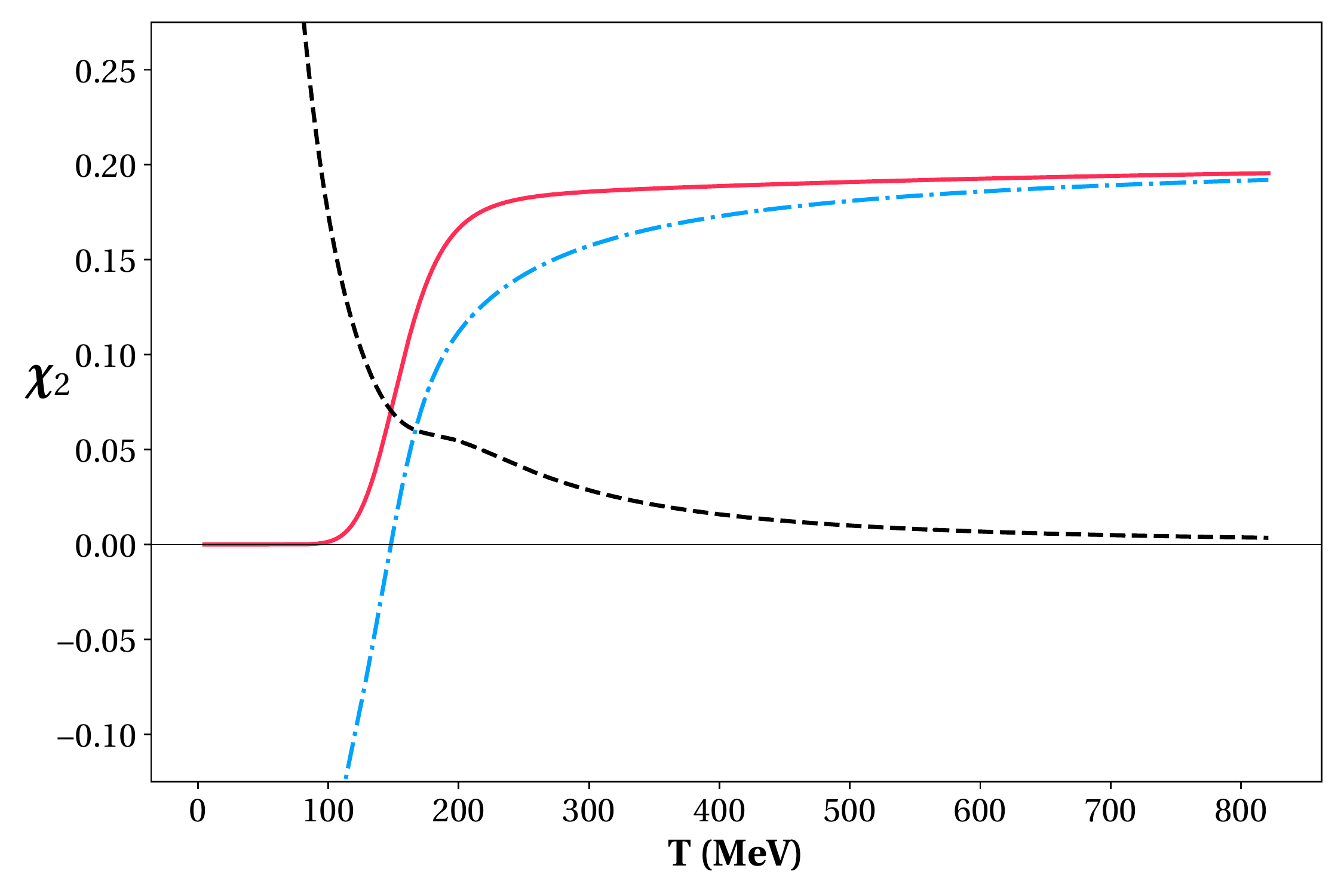} &
    \includegraphics[width=0.33\textwidth]{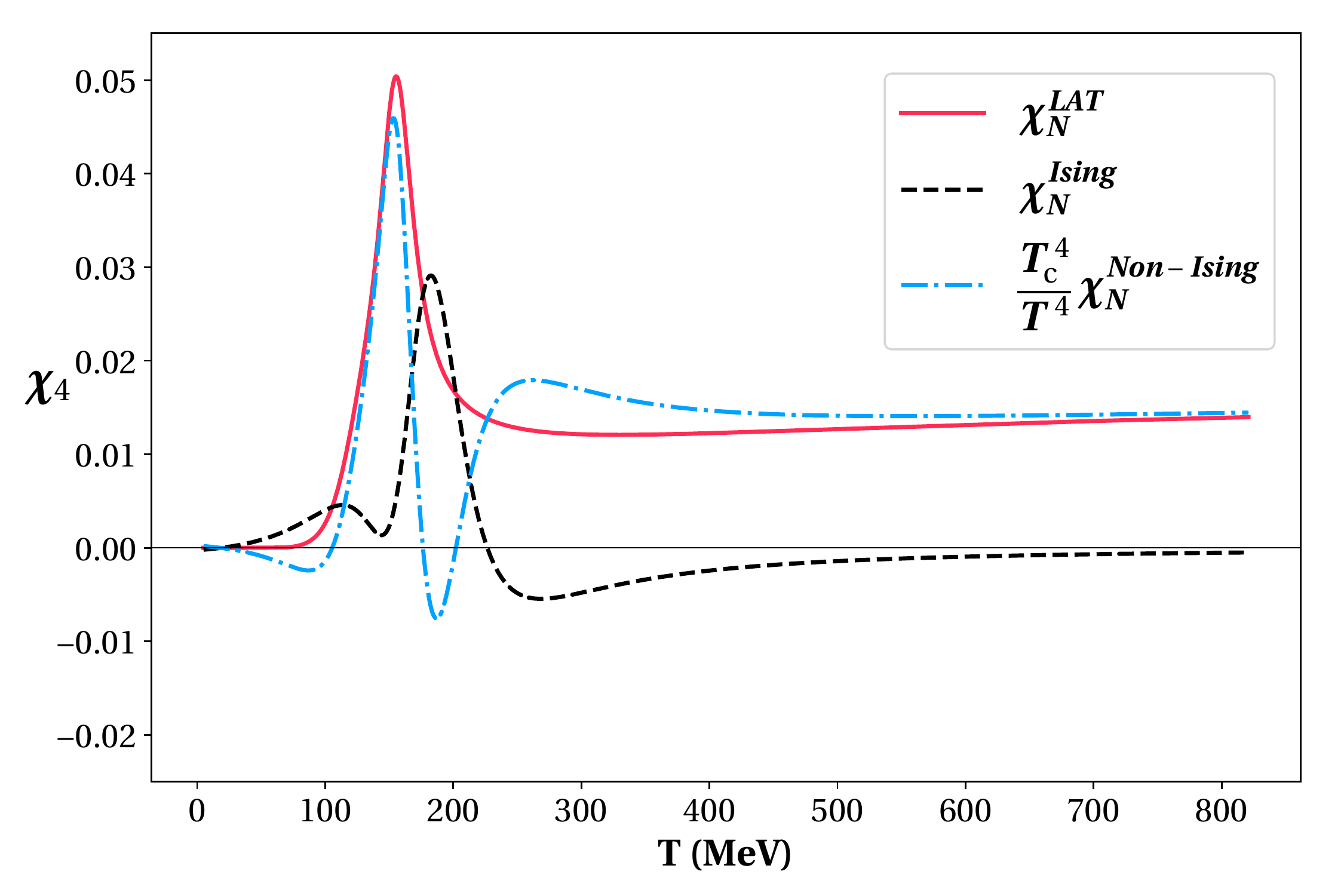}
    \end{tabular}
    \caption{Taylor expansion coefficients of the pressure up to $\mathcal{O}$($\mu_B^4$). The critical contributions from the Ising Model (black, dashed lines) are compared to the lattice QCD results (red, solid lines),  which determines the non-critical terms (blue, dot-dashed lines) as detailed in Eq.  (\ref{coeffmatch}). 
    }
    \label{fig:chis_all}
\end{figure*}
    \item Reconstruct the full pressure as the sum of critical and non-critical components
    \begin{equation} \label{fullpress}
    \begin{split}
        P(T,\mu_B)=T^4 \sum_n c_n^{\rm{Non-Ising}}(T) \left(\frac{\mu_B}{T} \right)^n
        \\
        + P_{\rm{crit}}^{\rm{QCD}}(T,\mu_B),
    \end{split}
    \end{equation}
    where $P_{\rm{crit}}^{\rm{QCD}}$ is the singular part of pressure that has been mapped onto the QCD phase diagram as described in steps 1 and 2.
    
    \item Merge the full pressure from step 5 with the pressure from the ideal HRG model, calculated using the SMASH hadronic list, at low temperatures in order to smoothen any non-physical artifacts of the Taylor expansion. We utilize the hyperbolic tangent for this smooth merging:
    \begin{equation} \label{eq:Pmerging}
    \begin{split}
        \frac{P_{\text{Final}}(T,\mu_B)}{T^4} = \frac{P(T,\mu_B)}{T^4} \frac{1}{2}
        \Big[1 + \tanh{\Big(\frac{T-T'(\mu_B)}{\Delta T}}\Big)\Big] \\
        + \frac{P_{HRG}(T,\mu_B)}{T^4} \frac{1}{2}
        \Big[1 - \tanh{\Big(\frac{T-T'(\mu_B)}{\Delta T}}\Big)\Big],
    \end{split}
\end{equation}
    where $T'(\mu_B)$ acts as the switching temperature and $\Delta T$ is the overlap region where both pieces contribute. The merging is performed along a line  parallel the QCD transition line with an overlap region of $\Delta T$=17 MeV.
    \item Calculate thermodynamic quantities as derivatives of the pressure from step 6.
\end{enumerate}
For a thorough description of this procedure, including an investigation of the parameter space and further discussion, we refer the reader to the original development of this EoS in Ref. \cite{Parotto:2018pwx}.  More details on the implementation of strangeness neutrality can be found in Ref. \cite{Karthein:2021nxe}.

We also calculate the correlation length as given in the 3D Ising model. For this quantity, we only provide the critical contribution as the authors are not aware of a calculation of the QCD correlation length on the lattice yet. We adopt the procedure from Refs. \cite{Brezin:1976pt, Berdnikov:1999ph, Nonaka:2004pg}, which follows Widom’s scaling form in terms of Ising model variables: 
\begin{equation} \label{corr_length}
    \begin{split}
        \xi^2(r,M) = f^2 |M|^{-2 \nu / \beta} g(x),
    \end{split}
\end{equation}
where $f$ is a constant with the dimension of length, which we set to 1 fm, $\nu$ = 0.63 is the correlation length critical exponent in the 3D Ising Model, $g(x)$ is the scaling function and the scaling parameter is $x$=$\frac{|r|}{|M|^{1/\beta}}$.  This scaling function takes two forms in the $\epsilon$-expansion and asymptotic formulations, which must be merged at an appropriate value of the scaling parameter, according to Ref. \cite{Nonaka:2004pg}. Further details on this procedure can be found in Ref. \cite{Karthein:2021nxe}.

\section{Results}
\begin{figure*}
    \centering
    \includegraphics[width=0.4\textwidth]{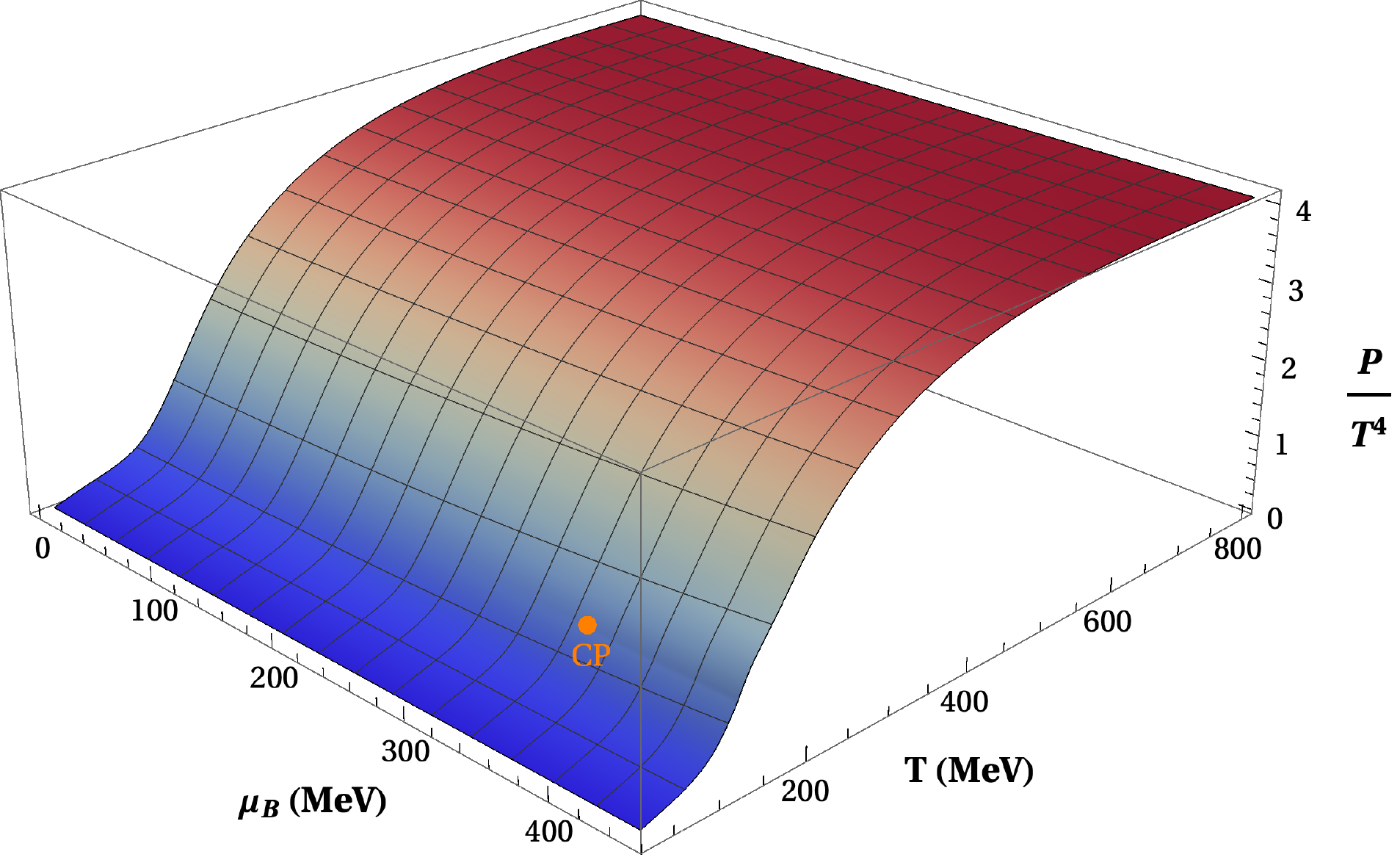}
    \includegraphics[width=0.49\textwidth]{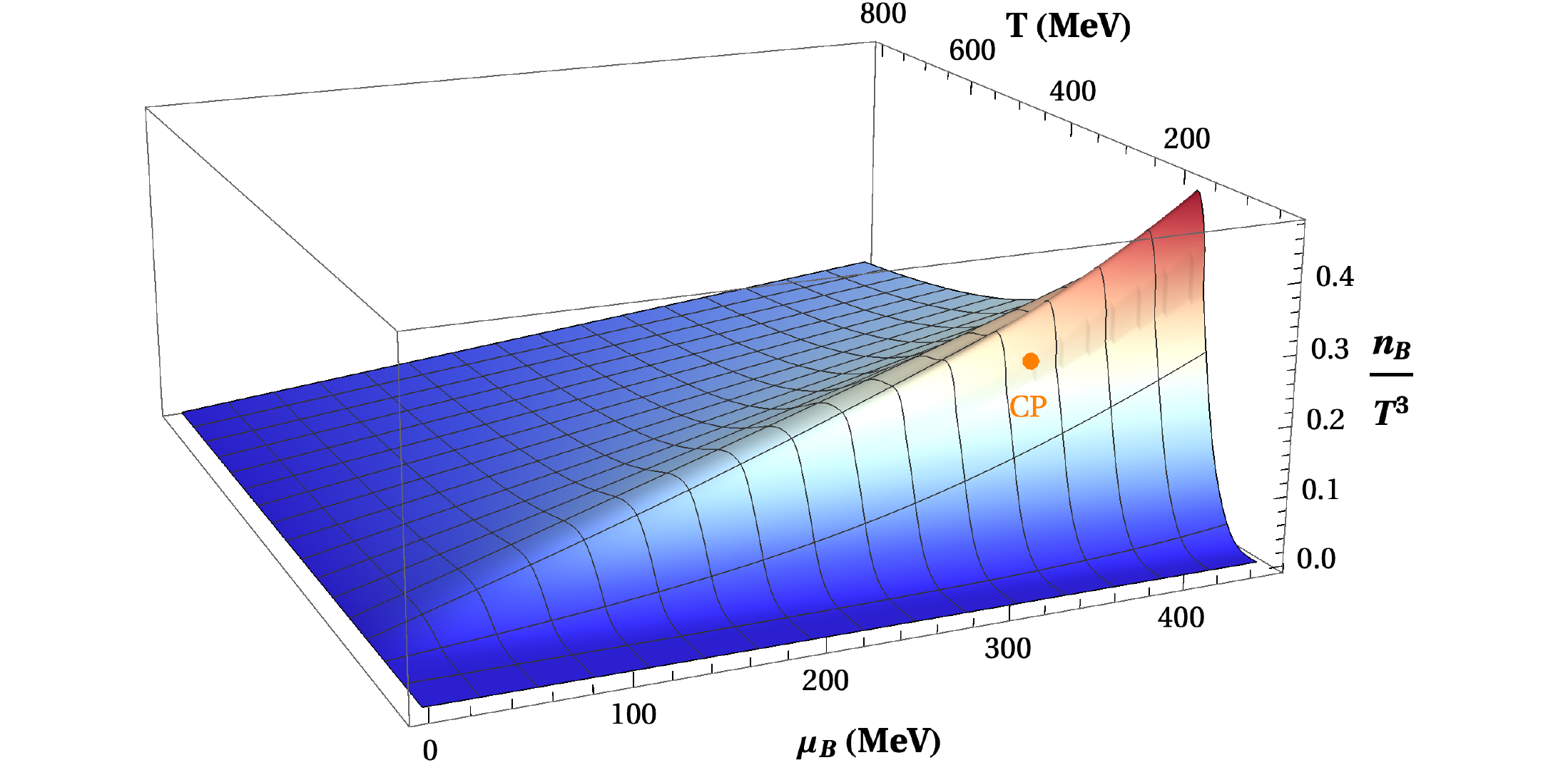}
    \caption{Left: The pressure for the choice of parameters as listed in Section \ref{method}. Right: The baryon density for the same choice of parameters.}
    \label{fig:press}
\end{figure*}

\begin{figure*}
    \centering
    \includegraphics[width=0.45\textwidth]{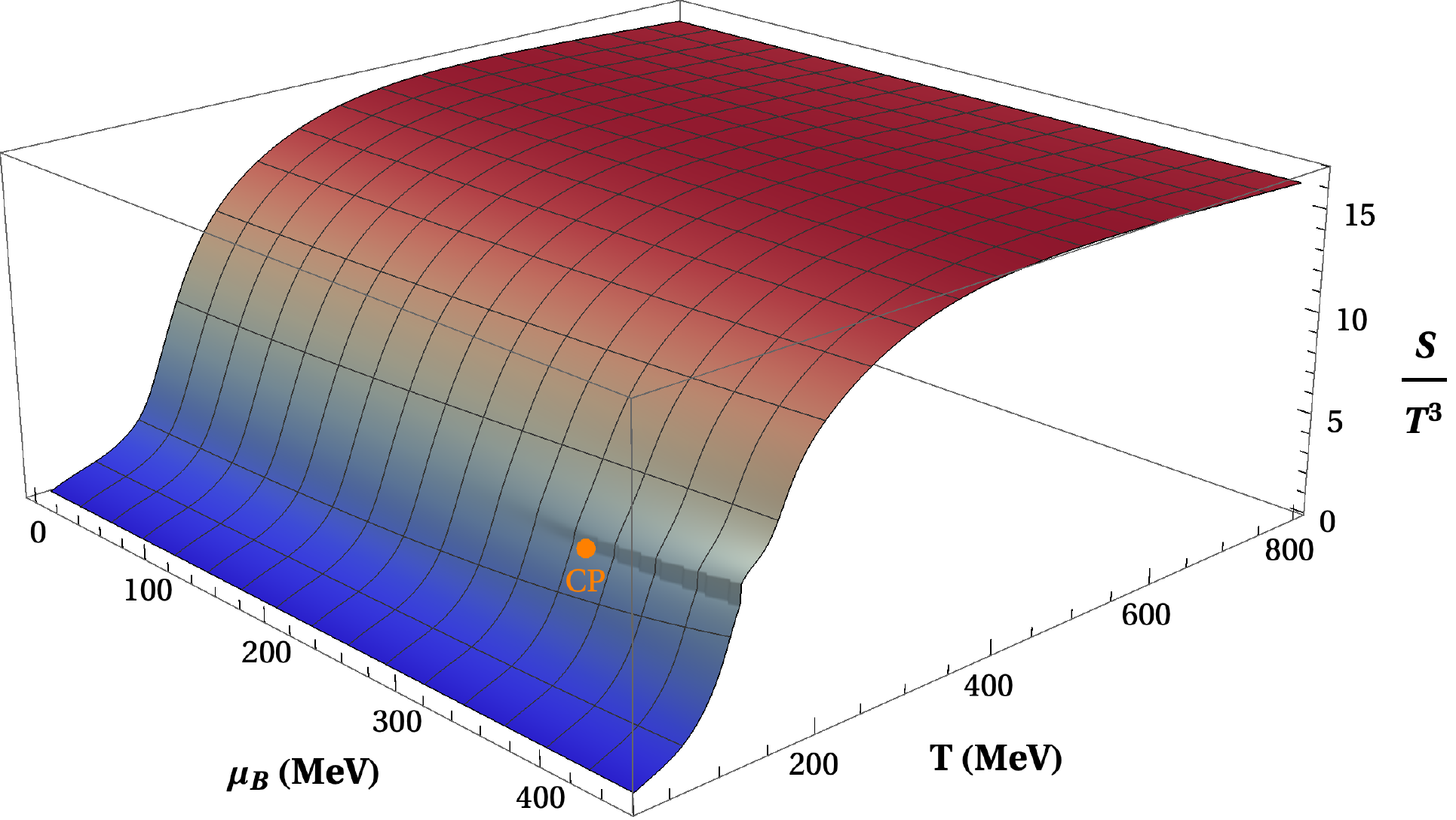}
    \includegraphics[width=0.45\textwidth]{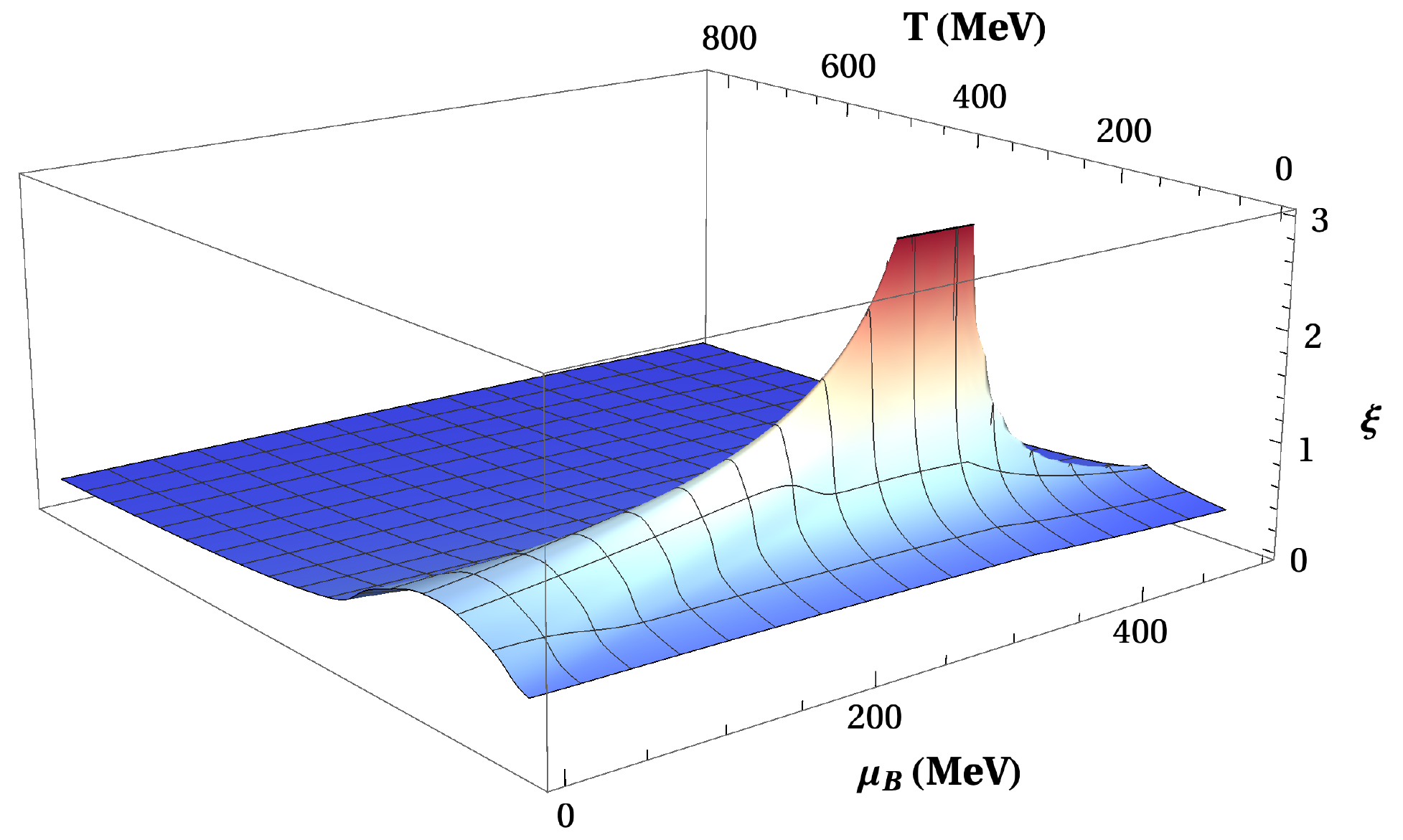}
    \caption{Left: The entropy for the choice of parameters as listed in Section \ref{method}. Right: The critical correlation length calculated within the 3D Ising model.}
    \label{fig:entr}
\end{figure*}

\begin{figure*}
    \centering
    \includegraphics[width=0.5\textwidth]{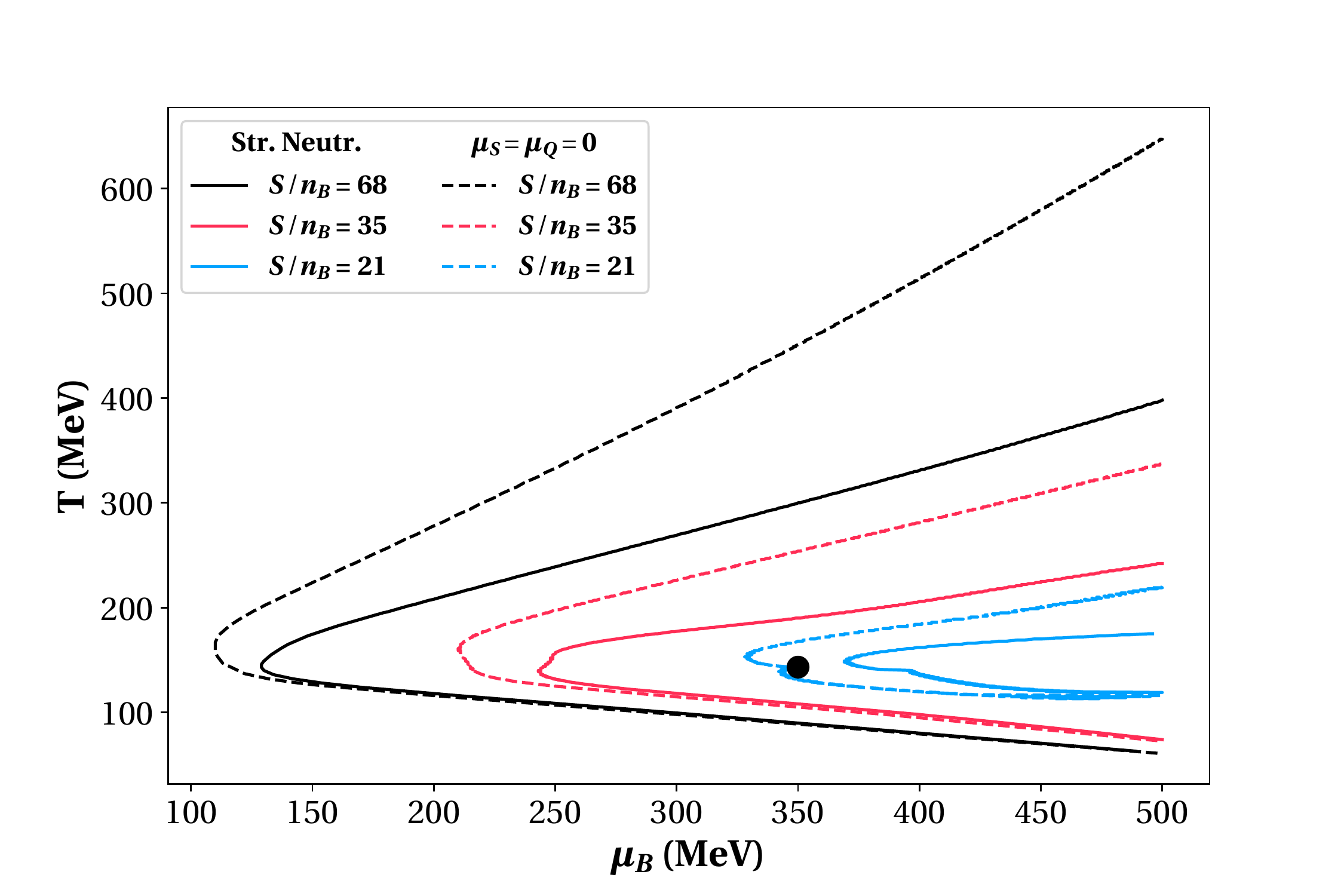}
    \caption{Isentropic trajectories in the QCD phase diagram that show the path of the system in a heavy-ion collision in the case of strangeness neutrality and vanishing $\mu_Q$ and $\mu_S$.}
    \label{fig:isen}
\end{figure*}

The results for the final thermodynamics are shown in Figs. \ref{fig:press} and \ref{fig:entr} along with the calculation of the correlation length.
The location of the critical point is indicated in each of the plots of the thermodynamics to guide the reader to the critical region. 
The pressure  itself is a smooth function of \{T,$\mu_B$\} in the crossover region and shows a slight kink for chemical potentials larger than $\mu_{B,c}$. 
The derivatives of the pressure help to reveal the features of the critical region,  as shown for the baryon density and entropy. 
In the right panel of Fig. \ref{fig:entr}, the critical contribution to the correlation length, calculated as described in section 2,  is shown in the QCD phase diagram.
The smoothly merged correlation length is uniform everywhere in the phase diagram, except in the critical region. 
Near the critical point the correlation length increases and then diverges at the critical point itself, following the expected scaling behavior.
We also calculated the isentropic trajectories in the phase diagram, as shown in Fig. \ref{fig:isen}, and note that these curves, in particular, show the importance of incorporating the constraints on the conserved charges present in HICs.

\section{Conclusions}
We provide an equation of state that obeys the experimental conserved charge conditions of strangeness neutrality and fixed baryon-number-to-electric-charge ratio to be used in hydrodynamic simulations of HICs. In addition to probing the slice of the 4D QCD phase diagram covered by the experiments, the equation of state can also be coupled to the SMASH hadronic afterburner because it exclusively includes the hadronic list from that framework \cite{Petersen:2018jag}.  Furthermore, we provide a calculation of the critical correlation length, which exhibits the expected scaling behavior and is important for the calculation of the critical scaling of transport coefficients near the critical point.

\section{Acknowledgments}
This material is based upon work supported by the
National Science Foundation under grants no. PHY1654219,  PHY-2116686 and OAC-2103680 and the National Science Foundation Graduate Research Fellowship Program under Grant No. DGE-1746047, and by the US-DOE
Nuclear Science Grant No. DE-SC0020633, and
within the framework of the Beam Energy Scan Topical (BEST) Collaboration. 
We also acknowledge the
support from the Center of Advanced Computing and
Data Systems at the University of Houston. P.P. acknowledges support by the DFG grant SFB/TR55.

\bibliographystyle{JHEP}
\bibliography{all}

\end{document}